\documentstyle[psfig,12pt,aaspp4]{article}
 
\def\hal{H$\alpha$}
\def\be{\begin{equation}}
\def\ee{\end{equation}}
\def\about{$\sim$}

\def\h{$h^{-1}$}
\def\HI{\ion{H}{1}}
\def\HII{\ion{H}{2}}
\def\NII{[\ion{N}{2}]}

\def\Ropt{$R_{\rm opt}$}
\def\Rcut{$R_{\rm cut}$}
\def\Dcl{$D_{\rm cl}$}

\def\Wopt{$W(R_{\rm opt})$}
\def\NRC{510}
\def\Nmem{429}
\def\Ntotmem{438} 
\def\Nfield{72}
\def\OG{$OG$}

\begin {document}
\title {Signatures of Galaxy-Cluster Interactions: Spiral Galaxy Rotation Curve Asymmetry, Shape, and Extent}
 
\author {Daniel A. Dale}\affil{IPAC, California Institute of Technology 100-22, Pasadena, CA 91125}
 
\author {Riccardo Giovanelli, Martha P. Haynes}\affil{Center for Radiophysics and Space Research and National Astronomy and Ionosphere Center, Cornell University, Ithaca, NY 14853}

\author {Eduardo Hardy}\affil{National Radio Astronomy Observatory, Casilla 36-D, Santiago, Chile}

\author {Luis E. Campusano}\affil{Observatorio Astron\'{o}mico Cerro Cal\'{a}n, Departamento de Astronom\'{\i}a, Universidad de Chile, Casilla 36-D, Santiago, Chile}

\begin{abstract}
The environmental dependencies of the characteristics of spiral galaxy rotation curves are studied in this work.  We use our large, homogeneously collected sample of \NRC\ cluster spiral galaxy rotation curves to test the claim that the shape of a galaxy's rotation curve strongly depends on its location within the cluster, and thus presumably on the strength of the local intracluster medium and on the frequency and strength of tidal interactions with the cluster and cluster galaxies.  Our data do not corroborate such a scenario, consistent with the fact that Tully-Fisher residuals are independent of galaxy location within the cluster; while the average late-type spiral galaxy shows more rise in the outer parts of its rotation curve than does the typical early-type spiral galaxy, there is no apparent trend for either subset with cluster environment.  We also investigate as a function of cluster environment rotation curve asymmetry and the radial distribution of \HII\ region tracers within galactic disks.  Mild trends with projected cluster-centric distance are observed:  (i) the (normalized) radial extent of optical line emission averaged over all spiral galaxy types shows a 4$\pm2$\% increase per Mpc of galaxy-cluster core separation, and (ii) rotation curve asymmetry falls by a factor of two between the inner and outer cluster for early-type spirals (a negligible decrease is found for late-type spirals).  Such trends are consistent with spiral disk perturbations or even the stripping of the diffuse, outermost gaseous regions within the disks as galaxies pass through the dense cluster cores.
\end{abstract}
 
 
\section {Introduction}

Observations over the past few decades have collectively established that the cluster environment plays an important role in the evolution of cluster galaxies.  From the well known excess of blue galaxies in intermediate redshift clusters (e.g. Butcher \& Oemler 1978; Couch et al. 1998) to morphological segregation (e.g. Dressler 1980; Dressler et al. 1997) and the existence of cD galaxies in many rich clusters, and from the \HI\ deficiency of cluster spirals (e.g. Haynes, Giovanelli \& Chincarini 1984; Solanes et al. 2000; Bravo-Alfaro et al. 2000) to observations of head-tail radio sources and ram pressure stripped spirals (e.g. Kenney \& Koopman 1999; Elmegreen et al. 2000), we can infer the effects the cluster environment has on cluster galaxies.  All of these phenomena are clues to the evolutionary history of galaxy clusters and many could well be linked to the presence of a rich intracluster medium which can be observed at X-ray wavelengths.  

Yet another line of evidence for the possible influence of the intracluster medium may be provided by the characteristics of spiral galaxy rotation curves within clusters.  Rubin, Ford \& Whitmore (1988) and Whitmore, Forbes \& Rubin (1988) first put forth the notion that spiral galaxy rotation curves exhibit a decline at the outer radii, in contrast with the asymptotically flat, or even rising, rotation curves that are typical of more peripheral cluster and field galaxies.  From that report it was inferred that the dark matter halos associated with the core galaxies have either been stripped due to galaxy-galaxy or galaxy-cluster interactions, or that the cluster environment inhibits
 the formation of dark matter halos.  Subsequent studies have produced conflicting conclusions.  Whereas Amram et al. (1993), using a sample of 19 galaxies, showed that the shape of the outer portions of rotation curves may mildly vary with the projected separation between galaxy and cluster core, Amram et al. (1996) used an expanded sample of 39 galaxies and found no trend.  This lack of a measurable trend was independently confirmed by Distefano et al. (1990) and Sperandio et al. (1995) who used samples of 15 and 32 (primarily Virgo) cluster galaxies, respectively.  Cluster-centric trends may be confused by variability in rotation curve shape coupled to the morphological segregation of cluster galaxies.  Adami et al. (1999) addressed this issue by separately analyzing the early and late-type spiral galaxies in the Amram et al. (1996) sample.  No trend was seen for Sb and earlier spirals, but a strong trend persisted for the sample of later-type spirals.  The impact of the cluster environment on spiral galaxy rotation curve shapes seems to be rather unclear, possibly due to the statistical uncertainty associated with small data samples.

The largest sample of cluster galaxy rotation curves obtained to date comes from the Tully-Fisher work carried out by Dale and collaborators (Dale et al. 1997, 1998, 1999; Dale \& Uson 2000).  In addition to  analyzing rotation curve shapes for those data, in this work we also study rotation curve asymmetry and extent as a function of the cluster environment.

We shall express distances in \h\ Mpc where the Hubble constant equals $H_\circ=100 h$ Mpc.


\section {The Data}
Our sample of \NRC\ optical rotation curves (of which \Ntotmem\ are from cluster members) derives from our compilation of Tully-Fisher data for 543 late-type disk galaxies spread throughout the fields of 53 Abell clusters in the redshift range $0.02 \lesssim z \lesssim 0.08$ (see Dale et al. 1997, 1998, 1999 and Dale \& Uson 2000).  The long-slit spectra were obtained at the Palomar 5 m and the CTIO 4 m telescopes.  The adopted observing strategy included a 300 second preliminary integration on each galaxy to verify cluster membership and get a quick estimate of the spectral line strength.  If the observation was deemed useful, a second integration between 600 and 3600 seconds was taken, depending on the strength of the emission lines in the outer disk regions.  The \hal\ line (6563 $\AA$) was primarily used to construct the rotation curves, but in \about 2\% of the cases the emission of a \NII\ line (6584 $\AA$) is stronger and extends to larger radii than that of the \hal\ emission; in these cases the \NII\ data were used.  In regions where heavy \hal\ absorption and emission are mixed, usually near the galactic center, we
fill in portions of the \hal\ rotation curve with data from the \NII\ rotation curve, if available.  We apply this \NII\ patch to approximately 10\% of the \hal\ rotation curves.  Rotational velocity widths $W$ were computed for each galaxy at radial distances referred to the optical radius \Ropt, where \Ropt\ is the distance along the major axis to the isophote containing 83\% of the $I$ band flux, and is reported by Persic, Salucci \& Stel (1996) and Giovanelli et al. (1997a, 2000) to be a useful radius at which to measure the velocity width of rotation curves.  In particular, it is found empirically that \Wopt\ recovers the \HI\ velocity width (measured at a level of 50\% of the profile horns).  Full details of the construction of the rotation curves and the measurement of their full velocity widths can be found in Dale et al. (1997, 1998, 1999).  

\begin{figure}[!ht]
\centerline{\psfig{figure=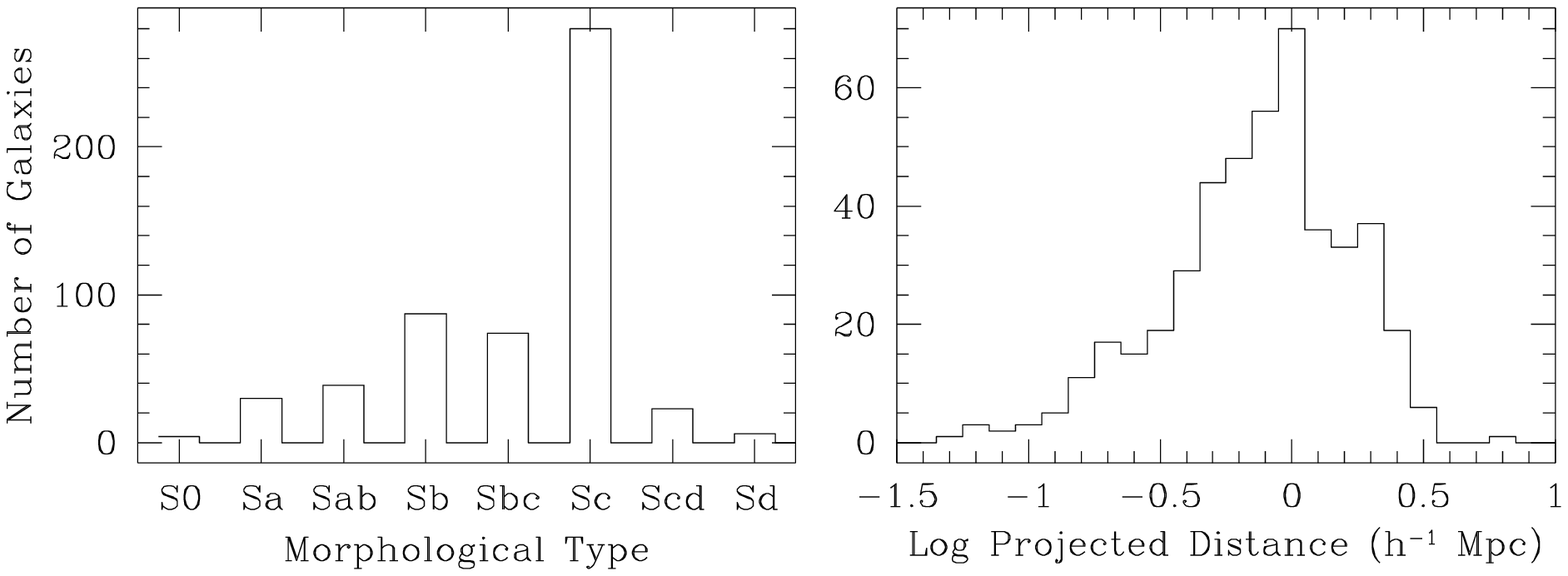,width=6in,bbllx=22pt,bblly=238pt,bburx=565pt,bbury=437pt}}
\caption[]
{\ Galaxy morphology distribution for the entire sample in the left panel and the projected cluster-centric distance distribution for cluster member galaxies in the right panel.}
\label{fig:hists}
\end{figure}

Since our goal is to investigate the influences of the cluster environment on galaxy rotation curves, it is useful to have a collection of galaxies of uniform morphology to minimize variations among rotation curves that depend on morphology.  Fortunately one constraint of our sample selection, the requirement that to recover high quality rotation curves we need the galaxies to be rich in ionized gas throughout their disks, encouraged a rather homogeneous sample comprised mainly of Sc types.  About 13\% of our galaxies are of a morphological type\footnote{Some 70 morphological classifications are rather uncertain due to the limited resolution of the images and the angular sizes of the smaller and more distant galaxies in our sample (see Dale et al. 1997, 1998, 1999 and Dale \& Uson 2000).} earlier than Sb, 16\% are type Sb, and 71\% are classified as Sbc or later.  The majority (52\%) of the galaxies are type Sc.  The left panel of Figure \ref{fig:hists} shows the morphological distribution of the galaxy sample.

Conversely, it is important to sample a wide range of cluster environments, from the innermost cores of rich X-ray clusters to the peripheries of less dense clusters and beyond.  In this way we can maximize the diversity of intracluster medium-interstellar medium interactions the galaxies are potentially experiencing.  The right panel of Figure \ref{fig:hists} displays the projected cluster-centric distances of our clusters.  In addition to these \Ntotmem\ cluster galaxies, we also have rotation curves from \Nfield\ field galaxies that lie in the foregrounds or backgrounds of the clusters.  Rotation curves from this subset are likely unaffected by intracluster media and thus can serve as a ``control sample'' to which we can compare data from cluster galaxies. 

\section {Results}
\label{sec:results}
\subsection {Rotation Curve Shape}
\begin{figure}[!ht]
\centerline{\psfig{figure=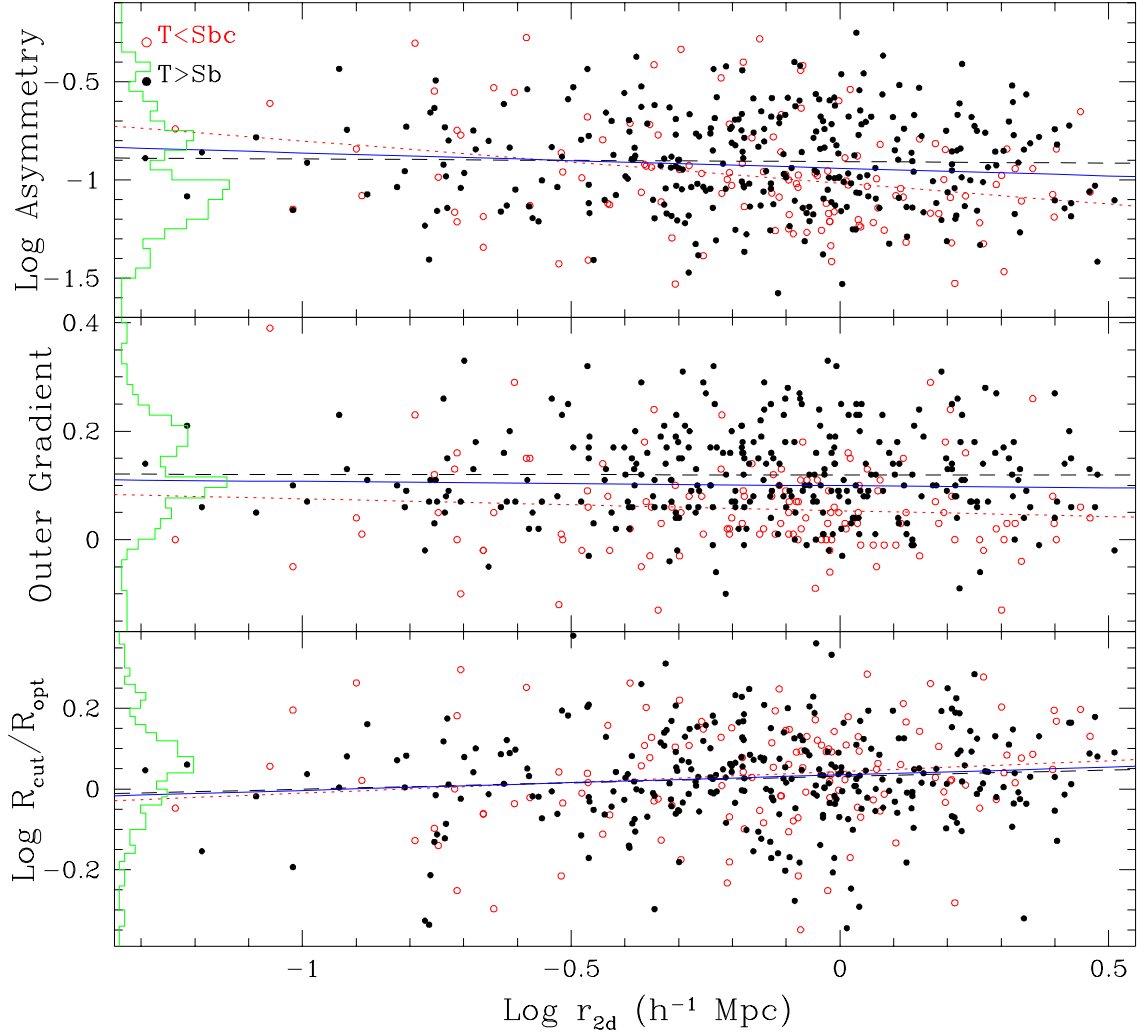,width=6in,bbllx=20pt,bblly=160pt,bburx=592pt,bbury=690pt}}
\caption[]
{\ {\bf Top panel}:  Rotation curve asymmetry as a function of projected cluster-centric distance (Equation \ref{eq:asymmetry}).  {\bf Middle panel}:  The outer gradient parameter for the rotation curves (Equation \ref{eq:outer_gradient}).  {\bf Bottom panel}:  The maximum radial extent of the observed emission line within the disk, normalized to the semi-major axis containing 83\% of the $I$ band flux.  Filled circles indicate Sb and earlier cluster galaxies while open circles show Sbc and later cluster galaxies.  The histograms at the left are the distributions for the foreground and background galaxies.  The solid lines are fits to the cluster galaxy data, and the dotted and dashed lines are separate fits to the early and late-type galaxy subsets (see Table 1).
}
\label{fig:og_vs_r2d}
\end{figure}

A simple but useful indication of the rotation curve shape in the outer disk region is the ``outer gradient'' parameter ($OG$), defined here as
\be
OG(\%) = 100 \times {V(R_{\rm opt}) - V(0.5R_{\rm opt}) \over V(R_{\rm opt})}
\label{eq:outer_gradient}
\ee  
where $V=V_{\rm rot}\sin i$ is the projected rotational velocity of the disk at inclination $i$.  In previous work, $R_{25}$, the 25 $B$ mag arcsec$^{-2}$ isophotal radius, was used instead of $R_{\rm opt}$ in the definition of $OG$.  We base our estimate of the optical radius on the $I$ band light profile since we only obtained $I$ band imaging for our sample (on average, though, $R_{\rm opt}\approx 1.45R_{25}$; Giovanelli et al. 1994; de Vaucouleurs et al. 1991).  Figure \ref{fig:og_vs_r2d} displays this parameter as a function of projected cluster-centric distance $r_{\rm 2d}$.  The average outer gradient for all cluster galaxies is $\langle OG \rangle=10.1 \pm 0.4$\%; a similar value is found for foreground and background objects ($10.6 \pm 1.1$\%).  Late-types typically show more rising rotation curves ($\langle OG_{\rm late} \rangle=11.9 \pm 0.5$\%) compared to early-types ($\langle OG_{\rm early} \rangle=5.7 \pm 0.8$\%).  The solid line overlayed is a fit to all the cluster galaxy data:
\be
OG(\%)=-0.8[\pm1.3](\log r_{\rm 2d}+\log h)+9.9[\pm0.5].
\ee
Such small slopes indicate a negligible dependence of rotation curve shape on cluster environment.  This may appear to contrast with earlier results from Whitmore, Forbes \& Rubin (1988), Amram et al. (1993), and Adami et al. (1999), and to be consistent with the findings of Distefano et al. (1990), Sperandio et al. (1995), and Amram et al. (1996).  However, all of these prior studies used far smaller samples of cluster galaxy rotation curves (see Table \ref{tab:slopes}); our large sample of rotation curves allows clearer conclusions to be drawn.   Our results are largely unchanged after culling the highly inclined ($i>75^\circ$) objects for which internal extinction effects may bias the shape of the rotation curve ($N_{i<75}=251$; \OG\ slope$=-2.0\pm1.6$\%).  Little change is expected since simulations have shown that only rotation curves from edge-on or nearly edge-on disks ($i\gtrsim87^\circ$) will be significantly affected by the effects of extinction (Bosma et al. 1992; Giovanelli et al. 2000; Matthews \& Wood 2000).

\begin{figure}[!ht]
\centerline{\psfig{figure=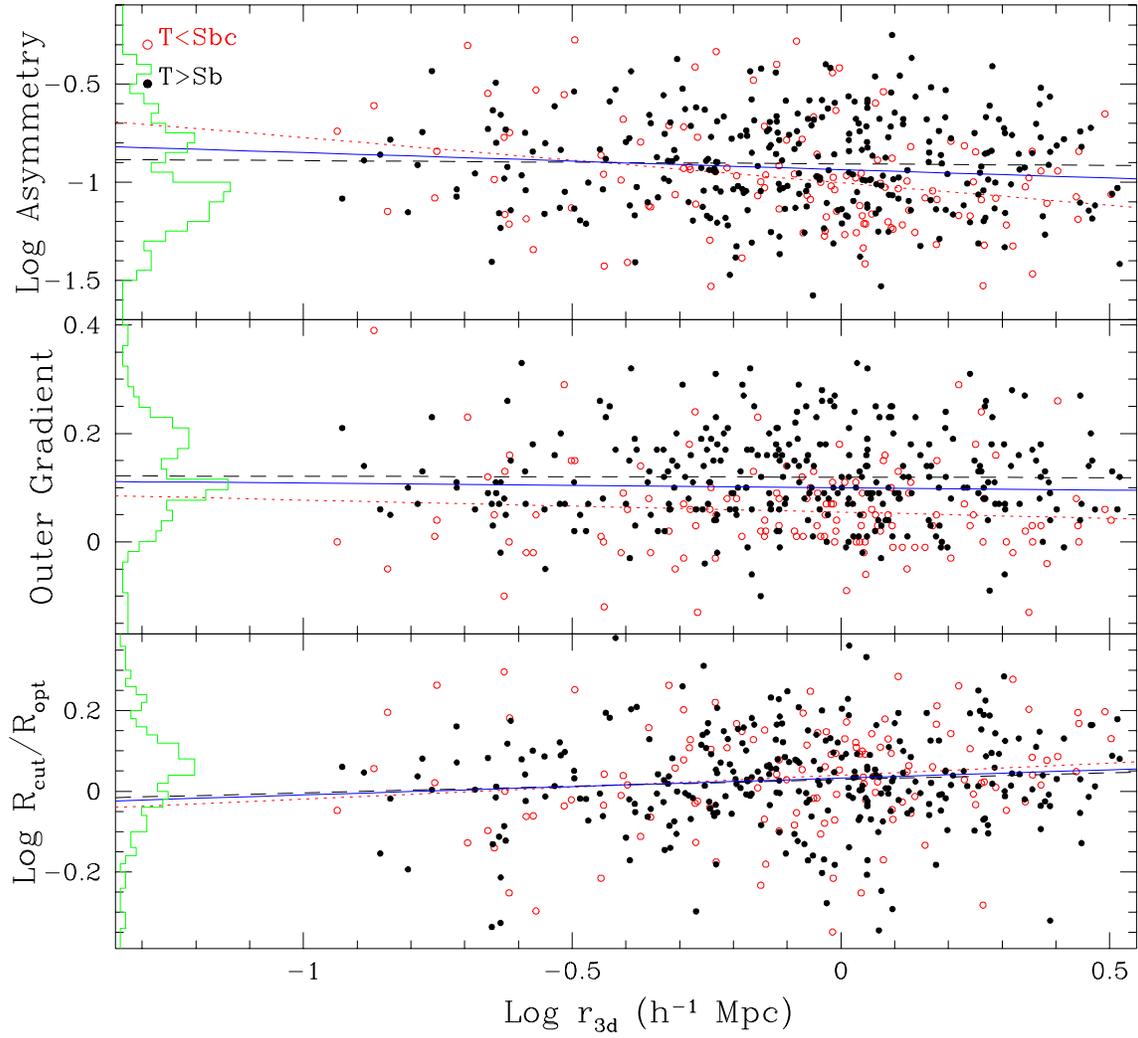,width=6in,bbllx=20pt,bblly=160pt,bburx=592pt,bbury=690pt}}
\caption[]
{\ Similar to Figure \ref{fig:og_vs_r2d} except that the cluster-centric distances have been statistically de-projected to three-dimensional distances.}
\label{fig:og_vs_r3d}
\end{figure}

Adami et al. (1999) tried to remove the effects of another bias by statistically de-projecting the observed, two-dimensional cluster-centric distances $r_{\rm 2d}$ to three-dimensional distances $r_{\rm 3d}$.  We have carried out a similar exercise assuming that galaxies are distributed within a cluster according to a King number density profile:
\be
n(r_{\rm 3d}) = n(0) (1+r_{\rm 3d}^2/r_{\rm core}^2)^{-3/2}
\label{eq:king}
\ee
where $r_{\rm core}$ is the cluster core radius.  We adopt the results of Adami, Biviano \& Mazure (1998), based on some 2000 galaxies in 40 clusters, for estimates of cluster core radii for different morphological types:
\begin{eqnarray}
r_{\rm core}&=&0.132 h^{-1} {\rm Mpc \;\;\;S0 \;galaxies}\\
            &=&0.159 h^{-1} {\rm Mpc \;\;\;Sa \;to \;Sbc \;galaxies} \nonumber \\
            &=&0.197 h^{-1} {\rm Mpc \;\;\;Sc \;and \;later}. \nonumber
\end{eqnarray}
For each cluster galaxy, we execute a series of Monte Carlo simulations to de-project cluster-centric distances.  A simulation is initiated by generating a random line-of-sight distance $D_{\rm los}$ such that $\Delta D_{\rm los} = |D_{\rm los} - D_{\rm cl}| <5$ \h\ Mpc and $r_{\rm 3d}=\sqrt{r_{\rm 2d}^2+\Delta D_{\rm los}^2}$ (i.e. $D_{\rm los}$ and \Dcl\ are the observer-galaxy and observer-cluster core separations, respectively).  This trial three-dimensional distance is kept only if it is likely: a second random number from the interval [0,1] must be less than $n(r_{\rm 3d})/n(0)$ given by Equation \ref{eq:king}.  This process is repeated until a list of $10^3$ $r_{\rm 3d}$ is produced.  The mean or median of those values is assumed to be statistically representative of the galaxy's de-projected distance, and the standard or absolute deviations are possible estimates of its uncertainty (Figure \ref{fig:og_vs_r3d} uses the median de-projected distance).  There is little change in the outer gradient trends after adopting these statistically de-projected distances (see Table 1 and Figure \ref{fig:og_vs_r3d}), and the results are nearly invariant with respect to reasonable variations of $r_{\rm core}$.

\subsection {Rotation Curve Extent}
We obtained our rotation curve database with the aim of measuring Tully-Fisher distances to our sample of galaxy clusters.  Since good data are needed in the outermost portions of rotation curves to reliably extract disk rotational velocities, we were especially careful to integrate long enough on each source to ensure high signal-to-noise spectral data for the outer disk regions.  Thus, our sample of rotation curves also provides a rich source for exploring the dependence of emission-line extent on environment.  Using a sample of 89 Virgo cluster galaxies, Rubin, Waterman \& Kenney (1999) find that the extent of measured emission \Rcut\ is a function of morphological type, with $R_{\rm cut}/R_{25}$ running from 0.3 for S0 galaxies to 0.7 for Sb and Sc types.  For our sample, the average radial extent of the rotation curves for all cluster galaxies is $\langle R_{\rm cut}/R_{\rm opt} \rangle=1.11 \pm 0.02$; similar values are found for early-type spirals, late-type spirals, and foreground and background spirals.  The reason we see little change with morphological type probably partly stems from our decision to preferentially observe sources with strong extended emission, in order to promote reliable rotational velocity extraction.  Nevertheless, there may still be a trend with cluster environment.  For example, though Rubin, Waterman \& Kenney (1999) observe no dependence with projected distance from the Virgo cluster central galaxy M87 for E and S0 galaxies, they do find a trend for later type galaxies (no quantitative result is given).  They claim this result is analogous to the well-known environmental dependence of Virgo \HI\ deficiency, where inner-cluster galaxies are more likely to have truncated neutral hydrogen disks (e.g. Giovanelli \& Haynes 1985; Haynes \& Giovanelli 1986).  The bottom panels of Figures \ref{fig:og_vs_r2d} and \ref{fig:og_vs_r3d} show the radial extent of our rotation curves versus cluster-centric distance.  We see that optical emission lines are observed at greater radial distances within disk galaxies that are located further from the cluster cores.  Quantitatively, the trend for all cluster galaxies is 
\be
\log\left({R_{\rm cut} \over R_{\rm opt}}\right) = 0.04[\pm0.02](\log r_{\rm 2d}+\log h) + 0.03[\pm0.01],
\ee
which is displayed as the solid line fit to the data; this result is essentially independent of morphological type.  
Moreover, these results are insignificantly changed if we instead utilize $r_{\rm 3d}$.  Though the maximum radial extent in optical line emission appears to grow with increasing galaxy-cluster core separation, we remark that it is only a 2$\sigma$ result.

\subsection {Rotation Curve Asymmetry}
Kinematic ``disturbances'' or asymmetry in galaxy rotation curves has recently been studied by Rubin, Waterman \& Kenney (1999), Swaters et al. (1999), Kornreich et al. (2000), Beauvais \& Bothun (2000), and Kannappan \& Fabricant (2000).  A galaxy's rotation curve asymmetry presumably echoes its recent interaction history, be it through galaxy-galaxy interactions or galaxy-cluster interactions (e.g. Conselice \& Gallagher 1999), since we expect gravitationally disturbed disk velocity fields to regularize within a few rotation cycles ($\lesssim 1$ Gyr).  Interestingly, it appears that up to half of all galaxies show significant rotation curve asymmetries or lopsided \HI\ profiles, even if they are apparently isolated field galaxies (e.g. Swaters et al 1999; Haynes et al. 1998; Richter \& Sancisi 1994).  

To measure the global rotation curve asymmetry for our galaxies we normalize the total area between the kinematically-folded approaching and receding halves by the average area under the rotation curve:
\be
{\rm Asymmetry}={\sum {\vert \mid V(R)| - |V(-R) \vert \mid\over \sqrt{\sigma^2(R) + \sigma^2(-R)}} \over \case{1}{2} \sum {|V(R)|+|V(-R)| \over \sqrt{\sigma^2(R) + \sigma^2(-R)}}}
\label{eq:asymmetry}
\ee
where $\sigma(R)$ is the uncertainty in the data at radial position $R$.  Our sample of galaxies includes preferentially inclined disk systems ($i\gtrsim45^\circ$) which, as pointed out by Kornreich et al. (2000), implies that our asymmetry parameter is more sensitive to noncircular than nonplanar motions.  The average asymmetry for our cluster galaxies is similar to that for our sample of field (foreground and background) galaxies, $14.0\pm0.4$\% vs. $12.5\pm1.0$\%, but as may be expected, a good match with the mode of the field distribution is found only at cluster-centric distances greater than $\sim1.5$\h Mpc. In addition, though we agree with the Rubin, Waterman \& Kenney (1999) result that there is little difference in the overall average asymmetry for early and late-type galaxies, we find a decreasing trend as a function of cluster-centric distance for early-types.  Specifically, we find that 
\begin{eqnarray}
\label{eq:asymmetry}
{\rm Log(Asymmetry)}&=&-0.21[\pm0.07](\log r_{\rm 2d}+\log h) - 1.02[\pm0.03] \;\;\;T\leq{\rm Sbc}\\
                   &=&-0.01[\pm0.04](\log r_{\rm 2d}+\log h) - 0.91[\pm0.02] \;\;\;T>{\rm Sbc}\nonumber \\
                    &=&-0.08[\pm0.04](\log r_{\rm 2d}+\log h) - 0.94[\pm0.01] \;\;\;{\rm all \; types}\nonumber
\end{eqnarray}
The coefficients found when using $r_{\rm 3d}$ instead of $r_{\rm 2d}$ are the same as the above numbers to within 15\%.  For the range $-1.2<\log r_{\rm 2d}<0.5$, this implies that the asymmetry falls from 17 to 7\% for early-types and from 13 to 12\% for late-types, as we proceed from the innermost to the outermost cluster regions.  If we make the broad generalization that early-type spirals are more likely than late-type spirals to have cluster orbits with a large radial component (e.g. Ramirez \& de Souza 1998; Ramirez, de Souza \& Schade 2000), then early-type spirals will more often transit the cluster core and experience the effects of a denser intracluster medium and stronger galaxy-galaxy and galaxy-cluster tidal effects.  Our results imply that these effects are observable for the early-types still near the cluster cores, before enough time has passed for these galaxies to travel to the cluster peripheries and before a few disk rotations will re-regularize the disk velocity field.  However, Rubin, Waterman \& Kenney (1999) observe larger kinematic disturbances for Virgo cluster galaxies that are further from the central galaxy M 87.  They also notice a tighter velocity distribution for these galaxies, comparable to that observed for Virgo ellipticals.  Thus, they logically argue that, although galaxies with radial orbits are more likely to be perturbed by transiting near the cluster core, they are also more likely to spend the bulk of cluster crossings near apocenter.  This argument assumes that the rotation curve asymmetry induced by such perturbations is long lasting.  Their inference, as well as those extracted from the bottom panels in our Figures \ref{fig:og_vs_r2d} and \ref{fig:og_vs_r3d}, are to be considered statistically weak.

\section {Discussion}

As outlined earlier, the combined observational evidence from the past few decades points to the cluster environment as a significant factor in the evolution of cluster galaxies.  If the cluster environment likewise strongly affects spiral galaxy optical rotation curves, it would establish that the mass within the optical radius of cluster galaxies also depends on environment.  Using the largest available sample of cluster galaxy rotation curves we see no change in the shape of the rotation curve with cluster-centric distance, for either early or late-type spirals (though the outer parts of rotation curves for late-types show, on average, twice as much rise compared to that for early-types).  Our result is conceptually in agreement with the findings of Biviano et al. (1990), Giovanelli et al. (1997b), and Dale et al. (1999).  They used Tully-Fisher data from, respectively, 166 galaxies from 10 clusters, 555 galaxies from 24 clusters, and 441 galaxies from 52 clusters to show there is very little environmental dependence on, effectively, mass-to-light ratios.

On the other hand, we find some evidence suggesting that the ionized regions responsible for the optical emission lines utilized in this work (\hal\ 6563 $\AA$ and \NII\ 6584 $\AA$) typically extend to larger distances within galaxies located in the peripheries of clusters, with respect to galaxies closer to the cluster cores.  This is consistent with the scenario whereby galaxies passing near the cluster cores are more likely to be experiencing the effects of ram pressure stripping, cluster gravitational/tidal forces, or increased interactions with other galaxies due to the higher volume density of galaxies near the core.  Thus we would expect the extent of gaseous regions to be truncated within the disk.  The amplitude of the observed truncation is fairly small, however, at a level of 4$\pm2$\% per Mpc.  Perhaps the small amplitude of the effect and the large scatter in the data are influenced by galaxies that already have passed through or near the cluster core, and are now observed to lie closer to the cluster periphery.  This is plausible since ram pressure stripping is expected to occur on timescales an order of magnitude shorter than a typical cluster crossing time (e.g. Fujita \& Nagashima 1999).  However, the observed trend is probably biased low by our spectroscopy observing strategy: for each galaxy we performed a 300 second preliminary test integration to gauge spectral line strength throughout the disk and thus whether the galaxy was worth pursuing with a longer integration.  This process selects against galaxies with anemic and severely truncated rotation curves, and such objects tend to lie near cluster cores.

Why do we not see cluster-centric trends for Tully-Fisher residuals and the shape of rotation curves, yet we do observe a mild dependence for the extent of the line emission?  Observations of atomic and molecular gas may hint to the answer.  Even though up to 90\% of \HI\ gas is presumably stripped as a spiral galaxy passes through the cluster core (Roberts \& Haynes 1994; Adabi, Moore \& Bower 1999), this gas represents only a fraction of the total galaxy mass, and much of it is distributed beyond the extent of optical rotation curves.  Conversely, Boselli et al. (1997) show that the deficiency parameter for $^{12}$CO(J=1$\rightarrow$0), gas that is more centrally concentrated in disks than neutral hydrogen, does not differ for field and cluster galaxies, and is thus not a function of environment.  In other words, the more diffuse and peripheral gas in galaxies is easier to strip away upon interactions with other galaxies or the cluster core, whereas the inner gas remains relatively intact.  In this scenario, while the outermost gaseous regions (including \hal-emitting \HII\ regions) of galaxy disks may be stripped away after passing near the cluster core and thus show somewhat truncated optical rotation curves, the observed mass-to-light ratio does not change since little mass within \Ropt\ has been removed.

Why do we not see cluster-centric trends for the asymmetry in late-type spirals, yet we tentatively observe an environmental dependence for their rotation curve extent?  One possible answer is that it is far easier to remove the outermost, diffuse gas in spiral disks than it is to affect the entire rotation curve; the asymmetry parameter we have defined is a global asymmetry measure, whereas the extent of the optical rotation curve merely depends on the presence of the outermost \HII\ regions.  In contrast to the late-type spiral galaxies, our early-types exhibit greater rotation curve asymmetry near the cluster cores.  This trend is consistent with the notion that early-type spirals tend to have more radial orbits and are thus more easily perturbed or stripped, as evidenced by their more extreme \HI\ deficiency in clusters (Solanes et al. 2000).

Our results agree with some previous studies on the shape of the rotation curve and cluster environment, but they also disagree with the findings of Whitmore, Forbes \& Rubin (1988) and Adami et al. (1999).  Apart from the small number statistics in previous efforts, the discrepancies may lie in how the rotation curve shapes are derived.  For example, we use optical sizes defined by $I$ band data,  whereas the other studies use $R_{25}$, the 25 $B$ mag arcsec$^{-2}$ isophotal radius.  We thus use slightly different definitions for the outer gradient (our definition for the outer gradient parameter is approximately proportional to $V(0.7R_{25})-V(0.35R_{25})$ whereas the traditional definition is proportional to $V(0.8R_{25})-V(0.4R_{25})$).  A second possible area for concern is that some of the rotation curves stem from two-dimensional spectral mapping (e.g. Amram et al. 1996).  This approach can likely yield a different outer gradient measure since rotation curves derived from single-slit observations may not yield the highest (projected) velocity differences observed in the disk.  It would be surprising, though, if these differences would introduce any significant bias as a function of cluster-centric distance.

If ram pressure stripping is the primary player in interstellar-intracluster interactions, then all of these studies make the specious assumption that a projected cluster-centric distance, or even a statistically de-projected distance, is a valid indicator of the intracluster environment.  This may hold true within a single cluster, as cluster X-ray surface brightness profiles are often azimuthally symmetric (e.g. Jones \& Forman 1999).  Nevertheless, since the cluster X-ray luminosity function spans three orders of magnitude (e.g. Ebeling 1997), the density of the intracluster medium must also vary from cluster to cluster; it is clear that using the projected radius within a cluster is a poor substitute for estimating the intracluster medium strength, especially when data from a broad sample of clusters are lumped together.  It would be better to use the X-ray emission to pin down the density of the intracluster medium, as Magri et al. (1988) did for the environmental dependence of \HI\ deficiency, and to combine that with some estimate of a typical crossing speed to estimate the ram pressure.  Such an effort is in progress.  Thus far, archival X-ray data (ROSAT PSPC and Einstein IPC) exist for all but seven clusters in our 53 cluster sample, and Chandra and XMM observations are planned for several of our clusters.  As a quick check of the influence of cluster X-ray luminosity, we have repeated the analysis outlined in Section \ref{sec:results} for the 18 low X-ray luminosity ($L_{\rm X}<10^{44}$ ergs s$^{-1}$) clusters and for the 14 high X-ray luminosity ($L_{\rm X}>10^{44}$ ergs s$^{-1}$) clusters in our sample that were studied by Jones \& Forman (1999).  Most slopes and zero points for both subsets are within 1$\sigma$ of the the numbers presented in Section \ref{sec:results} for all 53 clusters, and there is no systematic trend in the discrepancies with X-ray luminosity.

\section {Conclusions}
A sample of 510 spiral galaxy optical rotation curves is used to study the environmental influences of clusters.  In contrast to previous efforts, and using a database more than an order of magnitude larger in size, we find no significant trend in rotation curve outer slope as a function of cluster-centric distance.  Two additional parameters are also studied: rotation curve extent and asymmetry.  While late-type spirals show no systematic variation in kinematic asymmetry with cluster-centric distance, early-type spirals in cluster cores appear to be more perturbed than their counterparts in the cluster peripheries by a factor of two.  This result suggests that early-type spiral galaxies in clusters presumably have more radial orbits and thus more likely experience gravitationally-induced disturbances as they pass near the cluster cores, and that these distortions to their disk velocity fields diminish by the time a few rotations have completed and the galaxies have again reached the cluster outskirts.  Moreover, inner-cluster spirals typically show more extended optical line emission than outer-cluster spirals.  Similar to the increased \HI\ deficiency for spirals near cluster cores, this trend suggests that the outermost and less gravitationally bound diffuse gas in spiral galaxies is more easily stripped in or near cluster cores.
 
\acknowledgements 
Conversations with Alessandra Contursi, Gerry Neugebauer, and Philippe Amram improved the presentation of this work.  The results presented here are based on observations carried out at the Palomar Observatory (PO), at the Kitt Peak National Observatory (KPNO), and at the Cerro Tololo Inter-American Observatory (CTIO).  KPNO and CTIO are operated by Associated Universities for Research in Astronomy (AURA) under cooperative agreements with the National Science Foundation (NSF).  The Hale telescope at the PO is operated by the California Institute of Technology under a cooperative agreement with Cornell University and the Jet Propulsion Laboratory.  The National Radio Astronomy Observatory is a facility of NSF operated under cooperative agreement by AURA.  This research was supported by NSF grants AST94-20505 and AST96-17069 to RG and AST95-28960 to MPH.  LEC was partially supported by FONDECYT grant \#1970735.


\begin{deluxetable}{lrrll}
\small
\tablenum{1}
\def\p{$\pm$}
\def\a{\tablenotemark{a}}
\def\b{\tablenotemark{b}}
\def\c{\tablenotemark{c}}
\def\d{\tablenotemark{d}}

\tablewidth{0pt}
\tablecaption{Comparison of Rotation Curve ``Outer Gradient'' Trends with Cluster-Centric Distances}
\tablehead{
\colhead{Reference} & \colhead{$N$\a} & \colhead{Slope\b} & \colhead{Trend} & \colhead{Comment}
\nl
\colhead{} & \colhead{} & \colhead{\%} & \colhead{} & \colhead{}}
\startdata
Whitmore et al. 1988        &  16 &    43.2\p10.9 & strong     & \nl
Distefano et al. 1990 (D90) &  15 &       n/a~~~~ & negligible & \nl
Amram et al. 1993           &  19 &    12.6\p~5.4 & mild       & subset of A96 sample\nl
Sperandio et al. 1995       &  32 &       n/a~~~~ & negligible & some overlap with D90 sample\nl 
Amram et al. 1996 (A96)     &  39 &    0.1\p~3.4  & negligible & \nl
Adami et al. 1999           &  23 &   12.6\p~9.8  & mild       & subset of A96 sample \nl
                            &  23 &   35.8\p20.8\c& strong     & \nl
                            &   9 & $-$0.1\p~0.1\c& negligible & spirals earlier than Sbc\nl
                            &  14 &   38.6\p18.4\c& strong     & spirals later than Sb   \nl
Dale et al. 2001\d          &\Nmem& $-$0.8\p~1.3  & negligible & \nl
                            &\Nmem& $-$0.9\p~1.5\c& negligible & \nl
                            & 129 & $-$2.2\p~2.1  & negligible & spirals earlier than Sbc \nl
                            & 129 & $-$2.2\p~2.3\c& negligible & spirals earlier than Sbc \nl
                            & 300 & $-$0.1\p~1.4  & negligible & spirals later than Sb    \nl
                            & 300 & $-$0.2\p~1.6\c& negligible & spirals later than Sb    \nl
                            & 251 & $-$2.0\p~1.6  & negligible & disk inclination $\leq75^\circ$ \nl
                            & 251 & $-$2.1\p~1.7\c& negligible & disk inclination $\leq75^\circ$ \nl
\enddata
\tablenotetext{a}{The number of cluster member galaxies with measureable outer gradients.  Parent samples may be larger.}
\tablenotetext{b}{The slope computed from a linear fit to the outer gradient parameter as a function of the logarithmic cluster-centric distance in Mpc.  Though different Hubble constants are used in the different studies, this only impacts the {\it zero points} in the trends and not their {\it slopes}.}
\tablenotetext{c}{Using statistically deprojected three dimensional cluster-centric distances.}
\tablenotetext{d}{Outer gradient computed using galaxy diameters at $I$ band.}
\label{tab:slopes}
\end{deluxetable}
\normalsize

\end{document}